\documentclass[aps,pra,twocolumn,showpacs]{revtex4}
\usepackage{dcolumn}
\usepackage{bm}
\usepackage{graphicx}
\usepackage{amsmath}
\usepackage{latexsym}
\usepackage{amsfonts}
\usepackage{amssymb}
\usepackage{array}
\usepackage{epsfig}
\usepackage{times}
\usepackage{subfigure}
\usepackage{bbm}

\newcommand{\ket}[1]{\left\vert#1\right\rangle}
\newcommand{\modulus}[1]{\left\vert#1\right\vert}

\newcommand{\sprodsm}[2]{\left\langle#1\vert#2\right\rangle}

\newcommand{\projsm}[2]{\vert#1\rangle\langle#2\vert}

\newcommand{\sandsm}[3]{\langle#1\vert#2\vert#3\rangle}

\newcommand{\num}[1]{\hat{n}_{#1}}
\newcommand{\opa}[1]{\hat{a}_{#1}}
\newcommand{\nbar}{\bar{n}}
\newcommand{\uni}{{\cal \hat{U}}}
\newcommand{\unit}[1]{{\cal \hat{U}}_{#1}}
\newcommand{\pare}[1]{\left(#1\right)}
\newcommand{\be}{\begin{equation}}
\newcommand{\ee}{\end{equation}}
\newcommand{\ba}{\begin{array}}
\newcommand{\ea}{\end{array}}
\newcommand{\beqn}{\begin{eqnarray}}
\newcommand{\eeqn}{\end{eqnarray}}

\begin{document}
\title{Entanglement accumulation, retrieval, and concentration in cavity QED}

\author{Daniel Ballester}
\affiliation{School of Mathematics and Physics, Queen's University, Belfast BT7 1NN, United Kingdom}

\date{\today}

\begin{abstract}
The ability to accumulate and retrieve entanglement in the fields of two remote cavities with pairs of two-level atoms is discussed. It is shown that this transfer and retrieval can indeed be ideal with a resonant interaction. The case of initial non-maximally entangled atomic pairs is also considered. This leads to the possibility of concentrating entanglement into a single pair at the retrieval stage. A teleportation protocol based on the same setup is presented. This makes possible teleportation with built-in entanglement concentration.  
\end{abstract}


\maketitle

\section{Introduction}
The study of {\it entanglement} and non-classical correlations is of fundamental relevance to quantum information processing (QIP) research \cite{Horodecki}. It is behind some of the most profound puzzles in quantum mechanics such as the Einstein-Podolsky-Rosen paradox \cite{EPR} and non-locality \cite{Bell}. It has also been acknowledged that entanglement is a {\it resource} which turns out to be indispensable in various tasks of QIP, including quantum teleportation \cite{teleportation} and quantum algorithms \cite{nc}. This resource is very fragile because any interaction with the environment leading to decoherence would cause the irreversible loss of entanglement shared by different parties. This justifies the need for devising protocols and techniques with the aim of protecting the physical systems sharing entanglement from its surrounding environment, at least during the amount of time required for its subsequent usage \cite{Duan}.

Here the attention is focused on the transfer of entanglement from a bunch of entangled atomic pairs, which are assumed to be initially prepared off-line, to a system of two distant high-$Q$ cavities holding a single electromagnetic field mode per cavity. Different schemes for obtaining atomic Bell pairs as those required here can be found in Refs. \cite{Phoenix,Cabrillo,Bose,Cirac}. The process of entanglement transfer and its later retrieval (reciprocication) in these physical systems has been studied previously using the Jaynes-Cummings (JC) model \cite{JC} under resonant conditions \cite{reciproc} and also in its dispersive limit \cite{reciproc2}. It has also been shown how two distant cavities can accumulate entanglement after the successive passage of an atomic mediator through them \cite{accumulate}. This brings up, not only the possibility for transferring entanglement among qubits and continuous-variable systems \cite{Cirac,reciproc,reciproc2}, but also the ability of the latter to accumulate entanglement, in principle, indefinitely, which could be further exploited from a technological viewpoint.

There have been proposals of protocols that permit to concentrate in a probabilistic manner the level of entanglement of a group of non-maximally entangled parties \cite{Bennet}, by means of the usage of local operations and classical communications (LOCC). In this paper we consider the case of partially entangled atomic pairs passing through two distant cavities. After entanglement is accumulated in the cavities by these partially entangled atoms, a fresh pair of atoms prepared in a separable state is sent through the respective cavities. We show that a new entangled pair, which possesses more entanglement than any of the initial pairs that were used, can be obtained by applying this retrieval process. The performance of this method for entanglement concentration is compared to other concentration protocols and it is concluded that it could become advantageous in cases when the state of the partially entangled pairs is unknown and their level of entanglement can differ from pair to pair.

Furthermore, the possibility of realizing the quantum teleportation protocol based on the same setup is detailed. The goal here is to study the conditions under which the Bell measurement can be realized through the atom-field interaction and their subsequent projective measurement. Quantum teleportation with the entangled resource provided by two initial non-maximally entangled atomic pairs is also discussed. It is seen that the method allows a fidelity which is larger than the one attainable with the entangled resource given by any of the two initial pairs. In other words, our teleportation scheme includes a built-in entanglement concentration protocol for the quantum channel. Methods similar to those developed for quantum teleportation would allow us to perform entanglement swapping. Our scheme may be used to implement a quantum repeater \cite{Duan} which has been proposed for long-distance entanglement distribution on a quantum network of remotely located nodes. 

Finally, details of a possible experimental implementation of the setup and protocols devised here is given. The emphasis is put on the possibility to use the methods and technology developed in cavity quantum electrodynamics (cQED).

\section{Entanglement accumulation and retrieval}
\begin{figure}[t]
\centering{
\includegraphics[width=0.40\textwidth]{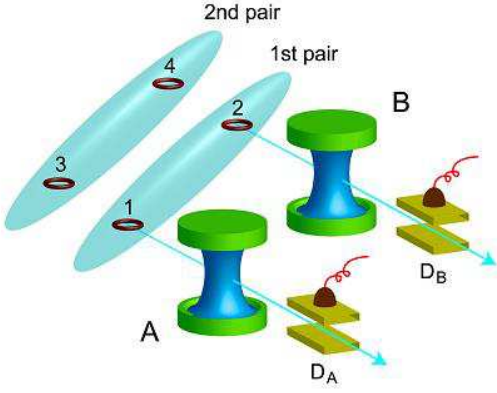} }
\caption{\label{fig:setup} Sketch of the setup proposed. Pairs of flying atomic qubits which are initially entangled pass sequentially through a pair of cavities initially prepared in a coherent state of large amplitude. After the detection of the state of the atoms, the entanglement in the atomic pairs is deterministically accumulated in the fields.}
\end{figure}

As depicted in Fig. \ref{fig:setup}, the physical system considered here consists of separable pairs of entangled two-level atoms together with two cavities held by Alice and Bob ($A$ and $B$). Each atom in the pair enters one of the cavities and interacts with a single cavity field mode for a period $t$. The interaction is assumed to be described by the resonant JC model \cite{JC}. During the interaction time $t$ the atoms-fields system evolves under the unitary operator $\uni = \unit{A}\otimes \unit{B}$. We can express the operator representing the partial evolution in each cavity, $\unit{j}$, $j=A,B$, using the atomic basis $\left\{ \ket{e}_j^a, \ket{g}_j^a \right\}$ of the atom interacting with the field in this cavity:
\be
\unit{j} = \left(
\ba{cc}
\unit{11}^{j} & \unit{12}^{j}\\
\unit{21}^{j} & \unit{22}^{j}\\
\ea \right) , \label{unitary}
\ee
where $\unit{\mu\nu}^{j}\equiv \unit{\mu\nu}^{j}(t)$, and \cite{PK}
\beqn
\unit{11}^{j}(t) = \cos\lambda t  \sqrt{\num{j}+1}   \quad\quad   \unit{12}^{j}(t) = -i \opa{j} \frac{\sin\lambda t \sqrt{\num{j}}}{\sqrt{\num{j}}}
\nonumber \\
\unit{21}^{j}(t) = -i \opa{j}^{\dag} \frac{\sin\lambda t \sqrt{\num{j}+1}}{\sqrt{\num{j}+1}}   \quad\quad   \unit{22}^{j}(t) = \cos\lambda t  \sqrt{\num{j}} 
\nonumber  
\eeqn
with coupling constant $\lambda$ and $\num{j}=\opa{j}^{\dag}\opa{j}$. The creation and annihilation operators of the photon field confined in cavity $j$ are denoted by $\opa{j}^{\dag}$ and $\opa{j}$.

Let us briefly review the entanglement deposition scheme proposed by Lee et al. \cite{reciproc}. Consider a pair of maximally entangled atoms in the state
\be
\ket{\psi(0)}_{AB}^{a1} = \frac{1}{\sqrt{2}} \pare{ \ket{e,g}_{AB}^{a1} + \ket{g,e}_{AB}^{a1} } , \label{psi0}
\ee
whereas the field in each cavity is initialized to a coherent state of real amplitude $\alpha$, $\ket{\alpha} = \exp(- \alpha ^ 2 /2) \sum_{n=0}^{\infty} \alpha^{n}/\sqrt{n!} \ket{n}$. Thus the total initial state is given by $\ket{\Psi(0)}_{AB}^{a1-f} = \ket{\psi(0)}_{AB}^{a1} \ket{\alpha}_{A}^f \ket{\alpha}_{B}^f$, where for simplicity we can also assume that $\alpha >0 $. After the system has evolved due to the interaction of atoms and field modes in each cavity, the state of the field modes can be post-selected as
\be
\ket{\psi(1)}_{AB}^{f}= \frac{1}{\sqrt{{\cal N}_{1}}} \pare{ \unit{21}^{A} \unit{22}^{B} + \unit{22}^{A} \unit{21}^{B} } \ket{\alpha}_{A}^f \ket{\alpha}_{B}^f   \label{psi1}
\ee
by measuring the state of the atoms in $\ket{g,g}_{AB}^{a1}$. Here ${\cal N}_{1}$ is the normalization constant. For small values of $\alpha$, the entanglement of this state shows a complicated pattern as a function of time, which is reminiscent of the JC model collapses and revivals of the population in the field mode \cite{reciproc}. In order to make the following calculations analytically tractable we will restrict ourselves to the case when $1\ll \lambda t \ll \alpha$. Under such conditions, it is legitimate to apply the continuous limit for the description of the coherent state in the cavities $\ket{\alpha}$, where the Poissonian distribution is approximated by the Gaussian distribution \cite{BR},
\be
\frac{\alpha^{2n}}{n!} \exp(-\alpha^{2}) \simeq \frac{1}{\sqrt{2\pi \alpha^{2}}} \exp(-x^{2}/2) , \label{gaussian}
\ee
and the sum over $n$ is replaced by integration over the variable $x=(n-\alpha^ 2)/\alpha$. It can be shown that with these approximations
$
\sandsm{\alpha}{ \unit{22}^{j \dag} \unit{21}^{j}}{\alpha} \simeq 0
$, and
$
\sandsm{\alpha}{ \unit{22}^{j \dag} \unit{22}^{j}}{\alpha} \simeq \sandsm{\alpha}{ \unit{21}^{j \dag} \unit{21}^{j}}{\alpha}
$,
therefore the state in the field modes (\ref{psi1}) is maximally entangled and the ebit of entanglement initially shared by the atoms in the pair has now been transferred to the cavity field modes \cite{reciproc}. Within this limit, which is asymptotically exact, the probability to project the state of the atoms into $\ket{g,g}_{AB}^{a1}$ is 1/4 \cite{reciproc}. Nonetheless, it is easy to see that this transfer of entanglement also occurs if the atoms leaving the cavities are projected into any of the 4 possible orthogonal states: $\left\{ \ket{g,g}_{AB}^{a1} , \ket{e,g}_{AB}^{a1} , \ket{g,e}_{AB}^{a1} , \ket{e,e}_{AB}^{a1}  \right\} $. Thus the transfer can be indeed performed {\it deterministically}.

After performing the entanglement accumulation within the cavity fields, we analyze the transfer of entanglement back to a pair of initially uncorrelated atoms. The initial state in the fields is the one given by $\ket{\psi(1)}_{AB}^{f}$ in Eq. (\ref{psi1}), whereas the pair of atoms is initialized to their ground state. After the interaction time $t $, the cavities are projected into the subspace spanned by coherent state of amplitude $\alpha$, leaving the atomic pair in the state $\ket{\psi(0)}_{AB}^{a1}$, i.e. if the process started with one pair of maximally entangled atoms, this state can be transferred to the cavities and later retrieved from them into a new pair of atoms, the probability of the retrieval being equal to 1/4 \cite{reciproc}. As with the case of entanglement accumulation, this retrieval step can be realized in a deterministic way, since it is easy to find a complete set of orthogonal projections for the state in Eq. (\ref{psi1}). Nevertheless, these other projective measurements might involve more tedious experimental tasks to be performed.

Whereas Lee et al. \cite{reciproc} focused only on the use of a pair of atoms to concentrate and retrieve entanglement in the cavity fields, here we consider a series of atomic pairs passing sequentially through the cavities. To study the capabilities of these cavity fields to accumulate more entanglement within their field modes, we can repeat the same process with a second pair of maximally entangled atoms prepared as in Eq. (\ref{psi0}), having the fields in the state given by Eq. (\ref{psi1}). After the atoms' passage through the cavities and projection into $\ket{g,g}_{AB}^{a2}$, the state in the fields becomes
\beqn
&& \ket{\psi'(1)}_{AB}^f  = \nonumber \\
&&   \left(  \unit{21}^{A}(t_2) \unit{21}^{A}(t_1)  \otimes \unit{22}^{B}(t_2) \unit{22}^{B}(t_1) \right. \nonumber \\
&& + \unit{22}^{A}(t_2) \unit{21}^{A}(t_1)  \otimes \unit{21}^{B}(t_2) \unit{22}^{B}(t_1)  \nonumber \\
&& + \unit{21}^{A}(t_2) \unit{22}^{A}(t_1)  \otimes \unit{22}^{B}(t_2) \unit{21}^{B}(t_1)  \nonumber \\
&& \left. + \unit{22}^{A}(t_2) \unit{22}^{A}(t_1)  \otimes \unit{21}^{B}(t_2) \unit{21}^{B}(t_1) \right)  \frac{\ket{\alpha}_{A}^f \ket{\alpha}_{B}^f }{\sqrt{{\cal N}_{1}'}} , \label{psi'1}
\eeqn
where we have distinguished between the interaction time of the first and second pairs, $t_1$, $t_2$. In order to achieve the aim of accumulating as much entanglement as it is available in the two pairs of atoms, under the same limiting conditions as before, it is possible to show that the selection of different interaction times for the two pairs is required. In particular, if $t_2 = 2 t_1$, then
$$
\sandsm{\alpha}{ \unit{2\zeta'}^{j \dag}(t_1) \unit{2\nu'}^{j \dag}(t_2) \unit{2\nu}^{j}(t_2) \unit{2\zeta}^{j}(t_1) }{\alpha} \simeq 0
$$
whenever $\nu\neq\nu'$ or $\zeta\neq\zeta'$, with the quantity
$
\sandsm{\alpha}{ \unit{2\zeta}^{j \dag}(t_1) \unit{2\nu}^{j \dag}(t_2) \unit{2\nu}^{j}(t_2) \unit{2\zeta}^{j}(t_1) }{\alpha} 
$
being unchanged for any $ \nu, \zeta(=1,2)$. This ensures that the state in Eq. (\ref{psi'1}) corresponds to a maximally entangled state of the field modes in a $2^2-$dimensional Hilbert space, i.e. the 2 ebits initially present in the two pairs of atoms have been accumulated in the fields of the cavities. Though the probability for measuring the state of the second pair of atoms in $\ket{g,g}_{AB}^{a2}$ is again 1/4, the statements on the transfer of entanglement to the cavity fields remain valid if the atoms are found in any of the other possible configurations. Certainly, this argument holds if the initial state chosen for the passage of the second atomic pair has been post-selected with a different atomic state in the first instance.


Assume now that after the passage of two pairs of atoms, the fields in the cavities are left with the state $\ket{\psi'(1)}_{AB}^f$ given in Eq. (\ref{psi'1}). If two fresh pairs of atoms prepared in their ground state are sent sequentially through the cavities, then after measuring both cavities in $\ket{\alpha}$, the state of the atomic pairs becomes
\beqn
&& \ket{\psi'(2)}_{AB}^{a1-a2} =  \nonumber \\ 
&&  \frac{1}{2} \pare{ \ket{e,g}_{AB}^{a2} +\ket{g,e}_{AB}^{a2} } \otimes  \pare{ \ket{e,g}_{AB}^{a1} +\ket{g,e}_{AB}^{a1} } , \quad\quad \label{psi'2}
\eeqn
i.e. the first and second pairs are left in a maximally entangled Bell states which do not share any entanglement between them. Thus, the 2 ebits initially accumulated into the field modes in the cavities are completely retrieved, and, in addition, the two pairs are still separable, as they were at the beginning. The probability for the post-selection of both cavities in $\ket{\alpha}$ is now 1/16.

\section{Entanglement concentration by accumulation}
Let us consider now the case in which a bunch of partially entangled atomic pairs (in a pure state) are provided,
\be
\ket{\varphi(0)}_{AB}^{a1} =  \lambda \ket{e,g}_{AB}^{a1} + \lambda' \ket{g,e}_{AB}^{a1}  , \label{phi0}
\ee
$ \lambda\geq 0$, $ \lambda'\equiv\sqrt{1-\lambda^ 2}$, where the degree of entanglement can vary among the different pairs. Taken as a single resource, the transfer of this state to the field modes in the cavities and its subsequent retrieval, following the process devised previously, is still viable. A more interesting situation arises if one wants to determine the state that can be obtained when two pairs of non-maximally entangled pairs (\ref{phi0}) are sent through the cavities, and only one single pair of fresh atoms is used at the retrieval stage.

\begin{figure}
\centering{
\subfigure[]{
\includegraphics[width=0.30\textwidth]{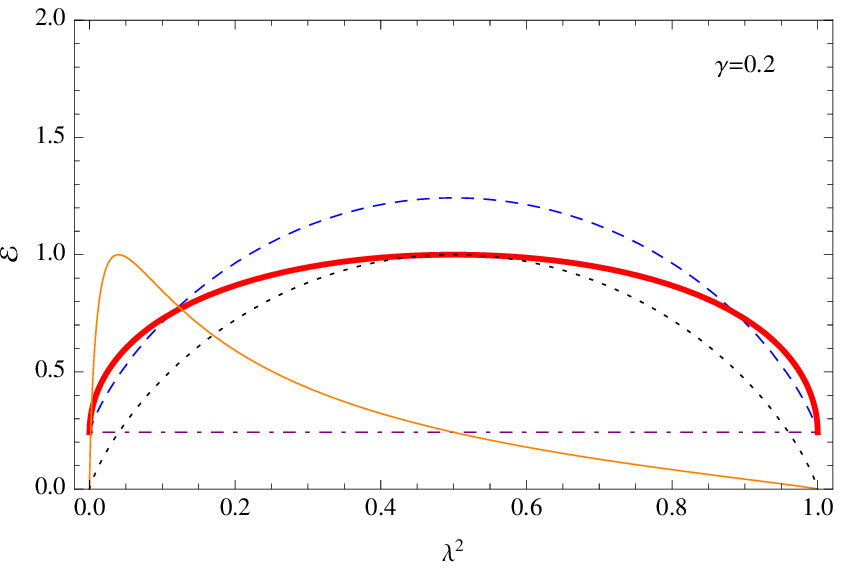}
\label{fig:fig-ent1}
}
\subfigure[]{
\includegraphics[width=0.30\textwidth]{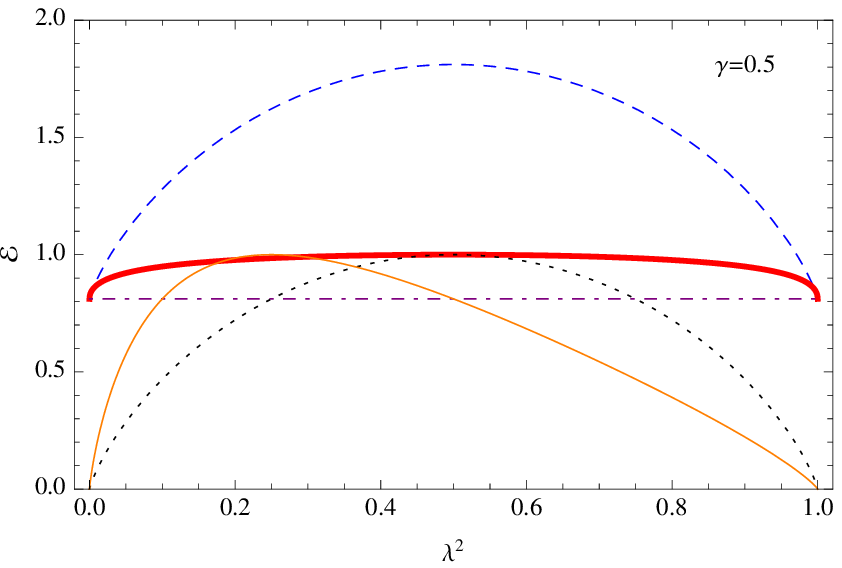}
\label{fig:fig-ent2}
}
}
\caption{\label{fig:ent} Entanglement obtained for the state in Eq. (\ref{phi'2}) (solid thick) after the accumulation and retrieval processes. The entanglement accumulated in the cavity fields (dashed) equals the sum of the one of the 1st (dotted) and 2nd (dash-dotted) atomic pairs  for $\gamma=0.2$ (a) and $0.5$ (b). The solid line represents the entanglement obtained after the application of the Schmidt projection [$\gamma=0.2$ (a) and $0.5$ (b)] or the Procrustean method [assuming that $\cos\phi=\gamma$ is chosen for the polarization-dependent reflector]. Entanglement is calculated using the Von Neuman entropy.}
\end{figure}

\begin{figure}
\centering{
\includegraphics[width=0.30\textwidth]{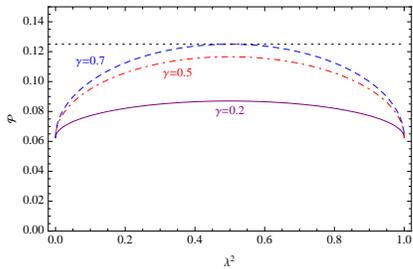} }
\caption{\label{fig:prob} Probabilities of obtaining the state in Eq. (\ref{phi'2}) after the accumulation and retrieval processes for different values of $\gamma = 0.2$ (solid), $0.5$ (dash-dotted), and $0.7$ (dashed). When the two atomic pairs are maximally entangled, the probability equals 1/8 (dotted).}
\end{figure}

It is clear that when non-equal partially entangled Bell pairs are used, the state left in the fields will take the form
\beqn
&& \ket{\varphi'(1)}_{AB}^f = 2\nonumber \\
&&  \left( \gamma \lambda  \unit{21}^{A}(t_2) \unit{21}^{A}(t_1)  \otimes \unit{22}^{B}(t_2) \unit{22}^{B}(t_1) \right. \nonumber \\
&& + \gamma' \lambda \unit{22}^{A}(t_2) \unit{21}^{A}(t_1)  \otimes \unit{21}^{B}(t_2) \unit{22}^{B}(t_1) \nonumber \\
&& + \gamma \lambda' \unit{21}^{A}(t_2) \unit{22}^{A}(t_1)  \otimes \unit{22}^{B}(t_2) \unit{21}^{B}(t_1)  \nonumber \\
&& \left. + \gamma' \lambda' \unit{22}^{A}(t_2) \unit{22}^{A}(t_1)  \otimes \unit{21}^{B}(t_2) \unit{21}^{B}(t_1) \right) \nonumber \\ && \frac{\ket{\alpha}_{A}^f \ket{\alpha}_{B}^f }{\sqrt{{\cal N}_{1}'}}  , \label{phi'1}
\eeqn
where $ \gamma\geq 0$,  $ \gamma'\equiv\sqrt{1-\gamma^ 2}$. This state contains the sum of the entanglement present in the initial atomic pairs. Next, a fresh pair of atoms, prepared in ground state passes through the cavities with an interaction time $t_1$, and after projecting both cavity fields into $\ket{\alpha}$, the atomic state will be post-selected as
\be
\ket{\varphi'(2)} _{AB}^{a1}= \vartheta \ket{e,e}_{AB}^{a1} + \vartheta' \ket{g,g}_{AB}^{a1} , \label{phi'2}
\ee
where $\vartheta = \pare{ \lambda \gamma' +  \lambda' \gamma } /{\cal N}$, $ \vartheta' = \pare{ \lambda \gamma +  \lambda' \gamma' }/ {\cal N}$, $ {\cal N} = \sqrt{ { \vartheta }^2 + { \vartheta'}^2 } $. The amount of entanglement retrieved in this state is depicted in Fig. \ref{fig:ent} for different values of initial entanglement. The key point here relies on the fact that the {\it retrieved entanglement} is always {\it higher than the entanglement present in any of the original pairs}. So this process can be applied as a protocol for {\it entanglement concentration} based on the accumulation of the entanglement contained in a bunch of partially entangled atomic pairs and its subsequent retrieval. The probability of the retrieval stage is shown in Fig. \ref{fig:prob}. It can be seen that it reaches 1/8 in case of having $\lambda = \gamma = 1/\sqrt{2}$. This result can be improved if we allow for a larger set of local unitary operations and projective measurements in the cavity fields, not just projection into $\ket{\alpha}$, as in Section IV.

It is instructive to compare the method for entanglement concentration of pure states presented here with other protocols devised for this purpose, namely, the Schmidt projection and the Procrustean method \cite{Bennet}. The former, which works provided that $n$ copies of the state to concentrate are given, is asymptotically efficient. The latter can generate a maximally entangled state out of a single copy of the partially entangled state, but the knowledge of the exact state to concentrate is required. Most theoretical proposals and experimental demonstration of entanglement concentration so far, have used an implementation of the Procrustean method, essentially due to practical reasons.

Nevertheless, the requirement of the exact knowledge of the state is rather strong under more realistic circumstances, where the interaction of entangled qubits with the environment is often not well understood. In order to obtain the information on the state of the entangled qubits, it is often the case that we have to destroy them as we measure them, causing the loss of qubits. Similarly, it is not pragmatic to require all the qubits to be in the same partially entangled state, as it is done in the Schmidt projection method. Therefore, in order to assess the performance of the concentration method by accumulation and compare it with the other two concentration schemes mentioned above, those two requirements will be relaxed. Similar assumptions have been used in Ref. \cite{SVK} in an entanglement-swapping-based protocol to concentrate non-maximally entangled states.

Take, in the first instance, the Schmidt projection protocol. Given two non-equal partially entangled qubits like (\ref{phi0}), the state of the pair obtained after applying this method \cite{Bennet} will be
\be
\ket{\varphi} _{AB}^{a1}= \frac{1}{\sqrt{\lambda^2 \gamma'^2 + \lambda'^2 \gamma^2 }} \pare{\lambda \gamma'   \ket{e,g}_{AB}^{a1} +   \lambda' \gamma  \ket{g,e}_{AB}^{a1} } . \label{phi}
\ee
In Fig. \ref{fig:ent}, the entanglement of this pair is compared to the one of the state (\ref{phi'2}) for different values of $\lambda$ and $\gamma$.  When the entanglement of both initial pairs is small and their states are similar, the Schmidt projection gives rise to a highly entangled output pair, whose entanglement can be larger than the one obtained by the accumulation and retrieval. If the pairs are initially highly entangled and are in a similar state, both methods behave analogously. However, the performance of the accumulation and retrieval technique becomes increasingly favored if the initial pairs have different levels of entanglement. A striking case occurs when one pair is perfectly entangled and the other one is nearly disentangled: whereas the accumulation and retrieval method will give rise to a maximally entangled pair, the Schmidt projection will produce a disentangled one.

Similar conclusions can be drawn regarding the Procrustean method. Now consider that a bunch of non-maximally entangled qubits like (\ref{phi0}) are shared. The state is assumed to be unknown and $\lambda$ varies from pair to pair. Alice's spins are sent through a polarization-dependent reflector, which lets the down spins pass through unmodified, but reflects the spins up. Then Alice tells Bob her results, and Bob discards his spin if Alice did so \cite{Bennet}. If the pair has not been discarded, then its state becomes like $\ket{\varphi}_C^a$ in (\ref{phi}), but with $\gamma$ replaced by $\cos\phi$, which represents the degree of freedom of the reflector. If this reflector is set up (by chance) in such a way that $\cos\phi=\lambda$, then Eq. (\ref{phi}) is indeed a maximally entangled pair. The amount of entanglement that can be obtained with this method can also be seen in Fig. \ref{fig:ent} if one replaces $\gamma$ by $\cos\phi$. Clearly, the concentration to the maximally entangled pair is obtained when $\cos\phi=\lambda$ and there are many pairs whose entanglement will decrease after the process. 

Thus, the method proposed here for entanglement concentration by accumulation and retrieval can be advantageous with respect to the Schmidt projection and Procrustean protocols, when there is a level of uncertainty in the state of the partially entangled pairs. Whereas in the latter one cannot guarantee that the entanglement is not going to worsen, the former is characterized by producing a new pair whose entanglement is higher than any of the two initial pairs, provided that these can be cast in the form of Eq. (\ref{phi0}).

The selection of the interaction time for the second pair of atoms as $t_2=2 t_1$ is an important requirement in all schemes presented here, as it allows to tailor the interference pattern which guarantees the orthogonality among different states to built up a higher-dimensional space. Numerical estimates for the concentration protocol show that with a proper set of parameters $\alpha$, $t_1$, there is no significant change in the entanglement retrieved out of the cavities when $t_2$ is allowed to vary by about 10\% around the value $2 t_1$, although the actual state retrieved can change periodically at a higher frequency within this interval.

It is also possible to study the resilience of the method under nonideal circumstances, i.e. when the atomic qubits are in a mixed state. The analysis of the effect that the depolarizing channel, with the three archetypical qubit errors (bit flip, sign flip, bit plus sign flip), and the amplitude damping channel, due to atomic decay, have over the performance of the method shows that in spite of the possible loss of purity in some of those cases, the ability of the protocol to enhance the entanglement by concentration is still preserved for a limited amount of mixedness in the initial qubits.

\section{Teleportation of qudits}
Quantum teleportation \cite{teleportation} is arguably one of the most paradigmatic tasks in QIP, for which entanglement is an indispensable resource. In this Section an alternative realization of the protocol for teleportation of atomic qudits is proposed, which exploits  the setup already presented for the accumulation and retrieval of entanglement, and where the Bell measurement required in the original protocol \cite{teleportation} is substituted by the atom-field interaction followed by a projective measurement of their states. The results given are derived under the same conditions aforementioned, namely, when $1\ll \lambda t \ll \alpha$.

Following the original protocol proposed for qubits, assume Alice and Bob share a maximally entangled state $\ket{\psi(1)}_{AB}^f$, as given by (\ref{psi1}) with interaction time $t_1$, accumulated in their respective cavity fields. Alice is given a third atomic qubit prepared in an unknown state
\be
\ket{\psi}_{C}^{a} = a  \ket{g}_{C}^{a} + b \ket{e}_{C}^{a} , \label{psic}
\ee
with $ \modulus{a}^2 +  \modulus{b}^2 =1$. After letting this third atom interact with her cavity field mode for a period $t_1$, she projects the state of the atomic qubit into $\ket{e}$ and that of the field into $\ket{\alpha}$. Then Bob takes an atomic qubit prepared in its ground state, $\ket{\psi}_{D}^{a}=\ket{g}_{D}^{a}$, and lets it interact with his cavity field for the same amount of time. After projecting the state of the field again in $\ket{\alpha}$, his outgoing qubit ends up in the unknown state $\ket{\psi}_{D}^{a}=\ket{\psi}_{C}^{a}$. Thus it is possible to perform the quantum teleportation of the unknown state given to Alice by using the entanglement already accumulated in the cavity fields. The case just described accounts for the teleportation of the state (\ref{psic}) conditioned on the post-selection by Alice of her atom in $\ket{e}$ and the cavity field in $\ket{\alpha}$. For completeness it is interesting to provide with all the possible outcomes that Alice can measure and how those condition the state of Bob's cavity field. The basis selected by Alice for the projective measurement in the protocol will be formed by $\{ \ket{e}_{A}^{a}, \ket{g}_{A}^{a} \}$ for the atomic qubit and $ \{ \ket{0}_{A}^{f}, \ket{1}_{A}^{f}, \ket{2}_{A}^{f} \} \equiv \{ \ket{\alpha}_{A}^{f}, \sqrt{2} \hspace*{1mm} \unit{22} (t_2) \ket{\alpha} _{A}^{f} , \sqrt{2} \hspace*{1mm} \unit{21}^\dag (t_2) \ket{\alpha}_{A}^{f} \}$ for the field in her cavity, which gives rise to a set of six mutually exclusive ocurrences $\ket{\mu}\ket{\nu}$. Each of those will project the state of Bob's cavity field into a state of the form
\be
\ket{\delta}_{B}^{f}=\frac{1}{\sqrt{{\cal N}_\delta}} \pare{ a_{\mu\nu}  \unit{22} (t_1) + b_{\mu\nu}  \unit{21} (t_1) } \ket{\alpha}_B^{f}  , \label{stateB1}
\ee
whose coefficients are given in Table \ref{table1} (see Appendix). The probabilities for Alice measuring each of the six possible instances is 1/4 if $\ket{\nu}=\ket{\alpha}$, and 1/8 otherwise. Therefore, one can identify the set of orthogonal Bell states $\{ \ket{\Psi^\pm}_{AB} , \ket{\Phi^\pm}_{AB} \}$ in the original teleportation protocol for qubits \cite{teleportation} with a set of four different outcomes given in Table \ref{table1}, each of them occurring with probability 1/4.  Assume, for instance, that the target state $\ket{\delta}_{B}^{f}$ is the one obtained when Alice measures $\ket{e}_{A}^{a}\ket{0}_{A}^{f}$. After she tells Bob about her outcome, he needs to apply a local unitary operation [namely, identity ($\mathbbm{1}_B$), bit flip ($\unit{22}^B(t_1)\leftrightarrow\unit{21}^B(t_1)$), phase flip ($\unit{22}^B(t_1)\leftrightarrow-\unit{22}^B(t_1)$), or bit plus phase flip ($\unit{22}^B(t_1)\leftrightarrow \unit{21}^B(t_1)$, followed by $\unit{22}^B(t_1)\leftrightarrow -\unit{22}^B(t_1)$)], in order to bring the state in his cavity field into the same state, achieving the teleportation of the state $\ket{\psi}_C^a$ with unit fidelity. Although the complete protocol described here is deterministic, for a practical implementation it would be more interesting to focus on the case of post-selecting $\ket{e}_A^a\ket{\alpha}_A^f$, as it has been discussed previously.

This protocol can be applied straighforwardly to the teleportation of higher-dimensional qudits. Take, for instance, the state $\ket{\psi'(1)}_{AB}^f$ given in (\ref{psi'1}) which accounts up to 2 ebits, shared by Alice and Bob. Now the unknown state provided to Alice is $2^2-$dimensional and is physically enconded in the state of a pair of two-level atoms,
\be
\ket{\psi'}_C^a = a \ket{g,g}_C^a + b \ket{e,g}_C^a + c \ket{g,e}_C^a  + d \ket{e,e}_C^a    , \label{psi'c}
\ee
which is assumed to be normalized. For simplicity only one of the possible outcomes of Alice's measurement is treated here. Alice passes these two atoms through the cavity in sequence, letting them interact with the field for a time $t_1$ and $t_2=2t_1$. Then she measures  the atoms in excited state and the cavity in $\ket{\alpha}$. After receiving Alice's results, Bob takes a pair of fresh atoms in their ground state and lets them interact also sequentially as Alice did. After he projects the state of his cavity field into $\ket{\alpha}$, his pair of atoms will be in the state $\ket{\psi'}_C^a$.

\begin{table}[t]
\begin{center}
\begin{tabular}{l || c | c | c | c}
 &  $\ket{e}\ket{0}$ & $\ket{g}\ket{0}$ & $\ket{e}\ket{1}$ OR $\ket{g}\ket{2}$ & $\ket{e}\ket{2}$ OR $\ket{g}\ket{1}$  \\ \hline\hline
$a_{\mu\nu}$  &  $-a$ & $-b$ & $a$ & $b$ \\ 
$b_{\mu\nu}$  &  $b$ & $a$ & $b$ & $a$  \\
LO  &  $\mathbbm{1}_B$ & $\unit{22}^B\leftrightarrow\unit{21}^B$ & $\unit{22}^B\leftrightarrow-\unit{22}^B$ & $\unit{22}^B\leftrightarrow \unit{21}^B$  \\
  &   &   &   & \& $\unit{22}^B\leftrightarrow -\unit{22}^B$  \\
\end{tabular}
\end{center}
\caption{\label{table1} Coefficients $a_{\mu\nu}$, $b_{\mu\nu}$ of the state (\ref{stateB1}) depending on the post-selection made by Alice. Here $ \{ \ket{0} , \ket{1} , \ket{2}  \} \equiv \{ \ket{\alpha} , \sqrt{2} \hspace*{1mm} \unit{22} (t_2) \ket{\alpha}  , \sqrt{2} \hspace*{1mm} \unit{21}^\dag (t_2) \ket{\alpha}  \}$. Each of the four outcomes occurs with the same probability. LO indicates the local unitary operation that Bob will apply to his cavity field in each case. The interaction time for those operators is $t_1$.}
\end{table}

\begin{figure}[t]
\centering{
\includegraphics[width=0.30\textwidth]{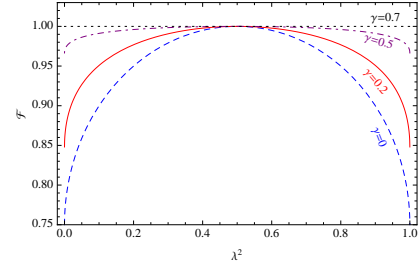} }
\caption{\label{fig:fidel} Average fidelities after the teleportation of the unknown state in Eq. (\ref{psic}) with two partially entangled cavity fields for different values of $\gamma = 0$ (dashed), $0.2$ (solid), $0.5$ (dash-dotted), and $0.7$ (dotted).}
\end{figure}

It is also possible to study the performance of this teleportation when the pairs of qubits provided are not maximally entangled. Assume now that the state accumulated in the cavities is the one given in (\ref{phi'1}), thus the entanglement shared by Alice and Bob could be smaller than 2 ebits. If Alice is given an atomic qubit in the unknown state $\ket{\psi}_C^a$, she lets this atom interact with the cavity field during an instant $t_1$. Now Alice could use the basis $\{ \ket{e}_{A}^{a}, \ket{g}_{A}^{a} \}$ for projecting the atomic qubit and $ \{ \ket{0}_{A}^{f}, \ket{1}_{A}^{f}, \ket{2}_{A}^{f} \} \equiv \{ \ket{\alpha}_{A}^{f}, \sqrt{2} \hspace*{1mm} \unit{22} (t_3) \ket{\alpha} _{A}^{f} , \sqrt{2} \hspace*{1mm} \unit{21}^\dag (t_3) \ket{\alpha}_{A}^{f} \}$ for the field with $t_3=2t_2$. Thus after measuring Alice's state in $\ket{\mu}\ket{\nu}$, the state in Bob's cavity field becomes
\beqn
\ket{\eta}_B^f =& \frac{ 1 }{\sqrt{{\cal N}_\eta}} \left( a_{\mu\nu} \unit{22}(t_2)\unit{21}(t_1) + b_{\mu\nu} \unit{21}(t_2)\unit{22}(t_1) \right. & \nonumber \\ +& \left. c_{\mu\nu} \unit{22}(t_2)\unit{22}(t_1)  + d_{\mu\nu} \unit{21}(t_2)\unit{21}(t_1) \right)  \ket{\alpha}_B^f    ,  \quad&  \label{stateB2}
\eeqn
where the coefficients are provided in Table \ref{table2}. For each of the four outcomes described here, Bob will apply a local unitary operation as before. To complete the protocol Bob needs to project his 4-dimensional state into the state of an atomic qubit. To this aim he can take a fresh atom in its excited state and let it interact with the cavity field for a period $t_1$. Finally, by measuring the state of the field in $\ket{\alpha}$, he will project the atom in the state
\be
\ket{\varphi'}_D^a = \frac{1}{\sqrt{{\cal N}_{\varphi _{1}'}}} \pare{a  \vartheta \ket{g}_D^a + b  \vartheta' \ket{e}_D^a} , \label{phi'd1}
\ee
if Alice measured either $\ket{e}_A^a\ket{0}_A^f$, $\ket{e}_A^a\ket{1}_A^f$ or $\ket{g}_A^a\ket{2}_A^f$, and
\be
\ket{\varphi'}_D^a =  \frac{1}{\sqrt{{\cal N}_{\varphi _{2}'}}} \pare{ a  \vartheta' \ket{g}_D^a + b  \vartheta \ket{e}_D^a } , \label{phi'd2}
\ee
if she measured either $\ket{g}_A^a\ket{0}_A^f$, $\ket{e}_A^a\ket{2}_A^f$ or $\ket{g}_A^a\ket{1}_A^f$. This result indeed coincides with the state that could have been obtained if the partially entangled state $\ket{\varphi'(2)}_C^a$ in (\ref{phi'2}) had been used as an entangled resource for the teleportation of $\ket{\psi}_C^a$ with the standard protocol \cite{teleportation}. Thus, the implementation proposed here enables us to achieve the teleportation of the unknown state as if it was done with a resource of larger entanglement. When using non-maximally entangled pairs, the fidelity of the state teleported ${\cal F}=   \modulus{{}_C^{}\hskip-0.15cm {}_{}^a\hskip-0.05cm\sprodsm{\psi}{\varphi'}_D^a}^2$ becomes smaller than 1 and is dependent on the state to teleport. The average fidelities obtained for various values of $\gamma$ are plotted in Fig. \ref{fig:fidel} as a function of $\lambda$. It must be pointed out that the protocol devised here is intrinsically probabilistic, as teleportation and entanglement concentration are done together. For that reason, the fidelity used does not account for the other possible outcomes in the fields' projective measurements, that is, it has to be divided by the overall probability to obtain the state (\ref{phi'2}) in the concentration protocol, when the same set of local unitary operations and projective measurements are allowed on the cavity fields. This definition is introduced just for the sake of comparison between the teleportation protocol with entanglement accumulation and the scheme formed by the concentration followed by the standard teleportation. The performance of the teleportation protocol with partially entangled pairs by accumulation of entanglement obviously shares similar features with its concentration counterpart: the {\it fidelity} obtained after the accumulation is {\it larger than the one that can be reached with any of the initial pairs}. It is instructive to recall that the fidelity attainable when an entirely classical resource is shared by Alice and Bob reaches an average value of 2/3 in the standard teleportation protocol for qubits \cite{Popescu}. In addition, the average fidelity compatible with local hidden variables theories is $\sim 0.87$ \cite{Gisin}.

\begin{table}[t]
\begin{center}
\begin{tabular}{l || c | c | c | c}
 &  $\ket{e}\ket{0}$ & $\ket{g}\ket{0}$ & $\ket{e}\ket{1}$ OR $\ket{g}\ket{2}$ & $\ket{e}\ket{2}$ OR $\ket{g}\ket{1}$  \\ \hline\hline
$a_{\mu\nu}$  &  $-a  \lambda' \gamma$ & $-b  \lambda' \gamma$ & $a  \lambda' \gamma$ & $b  \lambda' \gamma$ \\ 
$b_{\mu\nu}$  &  $a  \lambda \gamma'$ & $b  \lambda \gamma'$ & $a  \lambda \gamma'$ & $b  \lambda \gamma'$  \\
$c_{\mu\nu}$  &  $-b  \lambda \gamma$ & $-a  \lambda \gamma$ & $b  \lambda \gamma$ & $a  \lambda \gamma$ \\ 
$d_{\mu\nu}$  &  $b  \lambda' \gamma'$ & $a  \lambda' \gamma'$ & $b  \lambda' \gamma'$ & $a  \lambda' \gamma'$  \\
LO  &  $\mathbbm{1}_B$ & $\unit{22}^B\leftrightarrow\unit{21}^B$ & $\unit{22}^B\leftrightarrow-\unit{22}^B$ & $\unit{22}^B\leftrightarrow \unit{21}^B$  \\
  &   &   &   & \& $\unit{22}^B\leftrightarrow -\unit{22}^B$  \\
\end{tabular}
\end{center}
\caption{\label{table2} Coefficients $a_{\mu\nu}$, $b_{\mu\nu}$ of the state (\ref{stateB2}) depending on the post-selection made by Alice. Here $ \{ \ket{0}, \ket{1}, \ket{2} \} \equiv \{ \ket{\alpha}, \sqrt{2} \hspace*{1mm} \unit{22} (t_3) \ket{\alpha}  , \sqrt{2} \hspace*{1mm} \unit{21}^\dag (t_3) \ket{\alpha} \}$.  LO indicates the local unitary operation that Bob will apply to his cavity field in each case. The interaction time for those operators is $t_2$.}
\end{table}

Similar ideas could be used to implement an entanglement swapping protocol for entanglement distribution among remote nodes. In particular the ability to accumulate entanglement in the cavities leads to a protocol with built-in entanglement concentration capabilities, which can give practical advantages in some applications.

\section{Practical considerations}
This Section is devoted to discuss on the practical issues regarding the experimental implementation of the protocols developed in this paper. The experimental techniques and capabilities in cQED have been advancing enormously. Recent experimental realizations \cite{Haroche} make use of a Fabry-Perot resonator within the microwave regime built with reflecting superconducting mirrors. This type of cavities posseses a damping time of about $T_c=130$ ms, whereas its quality factor $Q$ can reach values of $4.2\times10^{10}$
and its finesse $4.6\times10^9$. This allows to limit the dissipation and decoherence in the field. The cavity is initialized by cooling it down to temperatures of $0.8$ K, which corresponds to an average of 0.05 residual photons. A classical pulse source is used to inject a microwave coherent field of amplitude $\alpha$.

As a two-level system, we can consider a Rb atom prepared in its circular Rydberg state with principal quantum number 50, which we take as the ground state $\ket{g}$. The cavity is tuned in resonance with the transition from this state to the one with principal quantum number 51, which acts as our excited state $\ket{e}$, at a frequency of 51.1 GHz. These atoms have a radiative lifetime of about 30 ms and a considerably large atomic dipole moment which permits to achieve a coupling of about $\lambda=2\pi\times50$ kHz. This lies well within the strong coupling regime. The atoms cross the cavity at a speed of about 250 m/s and are detected by a channeltron which can discern whether the atom is in its ground or excited state with a rate of 2 atoms/ms \cite{Deleglise}. The main resonator is usually surrounded by additional Ramsey zones (auxiliary microwave cavities) aiming to prepare the atoms in the required superposition state and can be used (connected to a microwave source) to form a Ramsey interferometer for diagnostic purposes. 

In addition, the protocols discussed rely on the implementation of a projective measurement of the cavity field into the coherent state at which they were initialized. This measurement could be realized by applying a second coherent pulse with the same source that is used to initialize the resonator, but now with opposite phase provoking the destructive interference of the field inside the cavity. This accounts for a displacement $\hat{D}(-\alpha)=\exp(-\alpha \hat{a}^\dag + \alpha^{*} \hat{a})$ of the state of the field \cite{BR}. Since $\projsm{\alpha}{\alpha}=\hat{D}(\alpha)\projsm{0}{0}\hat{D}(-\alpha)$, the projective measurement is reduced to the detection of vacuum after the displacement. The projection into vacuum can be done by the methods for measuring the photon number in the cavity field.

Along the previous discussion it was mentioned that the results derived are valid in the asymptotic limiting case of $1\ll \lambda t \ll \alpha$. It is important to note that in the simplest case of transferring and retrieving entanglement with a single pair of atoms, it suffices to assume that $1\ll \lambda t, \alpha$ \cite{reciproc}. Numerical estimates show that in practice, for the particular value $\lambda t = \pi$, $\alpha= 3$ gives a good approximation. For the rest of protocols involving higher number of pairs the amplitude required might be larger. For instance, a value of $\alpha\approx 10$ can be chosen if the error in the overlap \footnote{We refer here to the maximum error in the probability of finding the displaced field in vacuum state for the possible states in the protocol.} is allowed to vary up to about $8\%$ in the teleportation protocol with two non-maximally entangled qubits. Taking into account the value of the coupling $\lambda$, the interaction time required in the setup is short enough, of about $10^ {-2}$ ms, compared to the decoherence time $\tau \sim T_c / \nbar=T_c /\alpha^2$. In case of using a single pair of atoms, if the value $\alpha\approx 3$ is chosen then the repetition rate of 2 atoms/ms suffices for completing the protocol for transfer and retrieval of entanglement. However, if $\alpha\approx 10$ for the rest of the protocols, then it is clear that the decoherence time needs to increase in about one order of magnitude. Alternatively, the atomic rate could also be increased.

In order to implement a method for measuring the photon number in the cavity field after its displacement by $-\alpha$, an optimized procedure based on a quantum non-demolition scheme with an atomic interferometer and two-level atoms is described in detail in Ref. \cite{measure}. Following classical information theoretic arguments, the authors argue that this technique requires $\log_2 \Delta n$ atoms to converge to a Fock state, $\Delta n$ being the width of the initial distribution of photon number in the field. Explicit computations of these photon number probabilities, for instance in the cases of entanglement retrieval, show that $\Delta n\leq 20$ for the accumulation of 1 ebit with $\lambda t = \pi$, $\alpha=3$, while $\Delta n\leq 125$ for the accumulation of 2 ebit with $\alpha=10$.

Although the interest has been focused on the experimental implementation using cQED, it is also possible to test the same ideas in other physical setups. The most notable alternative is found in circuit QED, where the cavity is physically replaced by a superconducting coplanar waveguide transmission line resonator, whereas (stationary) superconducting Cooper pair boxes act as artificial two-level atoms \cite{Blais,Wallraff}. Circuit QED setups work in the optical regime and can reach very strong couplings, due to the large dipole moment. On the other hand, the lifetimes of both resonator and artificial atom are more limited, around 160 ns and 2 $\mu$s, respectively.

\section{Conclusions}
The entanglement of atomic two-level systems can be deterministically transferred to the fields of two resonators using a resonant interaction by tailoring the interaction time of each pair. This effect relies on the capacity of continuous-variable systems to accumulate entanglement and could be used as an entanglement storage device if the decoherence of the cavity fields is sufficiently small. Upon accumulation the entanglement can be either retrieved by fresh separable atoms, or exploited as a resource for various QIP tasks including teleportation and entanglement swapping to achieve the efficient distribution of entanglement in a quantum network. The system is robust against the presence of non-maximally entangled qubits and permits to transfer the accumulated entanglement into a small number of atomic pairs. 

\section{Acknowledgments}
The author thanks M. S. Kim, M. Paternostro, and C. Di Franco for encouraging discussions and insights. Funding from ESF is gratefully acknowledged.

\section*{Appendix: Teleportation of qubits with maximally entangled resources}
For the teleportation protocol, we assume that Alice and Bob cavities are in the maximally entangled state
\be
\ket{\phi}_{AB}^{f}= \sqrt{2} \pare{ \unit{21}^{A}(t_1) \unit{22}^{B}(t_1) + \unit{22}^{A}(t_1) \unit{21}^{B}(t_1) } \ket{\alpha}_{A}^f \ket{\alpha}_{B}^f   .\label{phi-ap}
\ee
Alice is given an atomic qubit in an unknown state $\ket{\psi}_C^a$ as in Eq. (\ref{psic}). After letting this atom to interact with her cavity field for a time $t_1$, the state of the joint system becomes
\beqn
\ket{\Upsilon}_{CAB}^{a-f}&=&\uni_A \otimes \mathbbm{1}_B \ket{\psi}_C^a \ket{\phi}_{AB}^{f} \nonumber \\
&=& \left( a \ket{g}_C^a \unit{22}^A(t_1) + a \ket{e}_C^a \unit{12}^A(t_1)   \right.  \nonumber \\
&& \left. b \ket{g}_C^a \unit{21}^A(t_1)  + b \ket{e}_C^a \unit{11}^A(t_1) \right)  \ket{\phi}_{AB}^{f}  ,  \label{ups1-ap}
\eeqn
which can be recast as
\beqn
\ket{\Upsilon}_{CAB}^{a-f}=\frac{1}{\sqrt2} \left( a \ket{g}_C^a \ket{\sigma}_{AB}^f + a \ket{e}_C^a \ket{\upsilon}_{AB}^f \right.  \nonumber \\
\left. b \ket{g}_C^a \ket{\upsilon}_{AB}^f + b \ket{e}_C^a \ket{\sigma}_{AB}^f \right), \label{ups2-ap}
\eeqn
with
\beqn
\ket{\sigma}_{AB}^f &=& \left\{ \left( \ket{0}_A^f  + \frac{\ket{1}_A^f }{\sqrt2} \right) \unit{21}^B(t_1) + \frac{\ket{2^\dag}_A^f}{\sqrt2}  \unit{22}^B(t_1) \right\} \ket{\alpha}_B^f  , \nonumber \\
\ket{\upsilon}_{AB}^f &=& \left\{ \left( - \ket{0}_A^f  + \frac{\ket{1}_A^f }{\sqrt2} \right) \unit{22}^B(t_1) + \frac{\ket{2^\dag}_A^f}{\sqrt2}  \unit{21}^B(t_1) \right\} \ket{\alpha}_B^f  , \nonumber 
\eeqn
$ \{ \ket{0}_{A}^{f}, \ket{1}_{A}^{f}, \ket{2^\dag}_{A}^{f} \} \equiv \{ \ket{\alpha}_{A}^{f}, \sqrt{2} \hspace*{1mm} \unit{22} (t_2) \ket{\alpha} _{A}^{f} , \sqrt{2} \hspace*{1mm} \unit{21} (t_2) \ket{\alpha}_{A}^{f} \}$, and $t_2=2 t_1$. The following approximations have been used above:
\beqn
\unit{22}(t_1)\unit{22}(t_1) \ket{0} &\simeq& \unit{11}(t_1)\unit{22}(t_1) \ket{0} \simeq \frac{1}{2}\ket{0} + \frac{1}{2\sqrt2}\ket{1} , \nonumber \\
\unit{12}(t_1)\unit{21}(t_1) \ket{0} &\simeq& \unit{21}(t_1)\unit{21}(t_1) \ket{0} \simeq - \frac{1}{2}\ket{0} + \frac{1}{2\sqrt2}\ket{1} , \nonumber\\
\unit{22}(t_1)\unit{21}(t_1) \ket{0} &\simeq& \unit{11}(t_1)\unit{21}(t_1) \ket{0} \simeq \unit{21}(t_1)\unit{22}(t_1) \ket{0} \simeq \nonumber\\
 && \unit{12}(t_1)\unit{22}(t_1) \ket{0} \simeq  \frac{1}{2\sqrt2}\ket{2^\dag} . \nonumber
\eeqn
These are valid in the asymptotic limit $1\ll \lambda t\ll \alpha$. From here one can easily obtained the values in Table \ref{table1}.


\end{document}